\documentstyle[12pt]{article} 
\begin{document} 

\begin{center} 
                {\large  Cherenkov gluons at RHIC and LHC}\\ 

                            I.M. Dremin\\

              {\it   Lebedev Physical Institute, Moscow }\\
\end{center} 

\begin{abstract} 
The coherent hadron production analogous to Cherenkov radiation of photons 
gives rise to the ring-like events. Being projected on the ring diameter 
they produce the two-bump structure recently observed for the away-side 
jets at RHIC. The position of the peaks and their height determine such
properties of the hadronic medium as its 
nuclear index of refraction, the parton density, the free path length and 
the energy loss of Cherenkov gluons. Beside comparatively low energy gluons 
observed at RHIC, there could be high energy gluons at LHC, related to the 
high energy region of positive real part of the forward scattering amplitude 
and possessing different characteristics. 
\end{abstract} 

Analogous to Cherenkov photons, the Cherenkov gluons \cite{1, 2, 3, 4, a} can 
be emitted in hadronic collisions provided the nuclear index of refraction 
$n$ exceeds 1. The partons moving in such nuclear medium would emit them. 
These gluons should be emitted at the cone surface with the cone angle 
$\theta $ in the {\bf rest system} of the {\bf infinite} medium defined by 
the relation 
\begin{equation} 
\cos \theta = \frac {1}{\beta n},   \label{cos} 
\end{equation} 
where $\beta $ is the ratio of the velocities of the parton-emitter and 
light which can be replaced by 1 for relativistic partons. 

{\bf Prediction 1.} According to Eq. (\ref{cos}) the ring-like two-dimensional
distribution of particles must be observed in the plane perpendicular to 
the momentum of the parton-emitter. 

{\bf Proposal 1.} Plot the one-dimensional pseudorapidity 
($\eta =-\ln \tan \theta /2$) distribution with trigger momentum as $z$-axis
neglecting the mismatch of trigger and away-side jets directions. 
It should have maximum at (\ref{cos}).

This plot is still unavailable at RHIC. RHIC experiments \cite{5, 6} have 
shown the two-bump structure of the azimuthal angle distribution (now with 
$z$-axis chosen along the collision axis) near the away-side jets. It 
results due to the one-dimensional projection of the ring on
the azimuthal plane. Ring's plane is perpendicular both to the trigger momentum
and to the plane in which momenta of colliding particles and trigger are placed.
It is clear that projection of a ring on its diameter in the azimuthal plane 
is not the best one to reveal
its properties. The proposal 1 uses better (circular) projection of the ring. 
The shapes of two- and three-particle correlations studied at RHIC \cite{pr,zh}
are its less direct indications although they have the ring-like structure 
themselves.

 From the distance between the peaks defined in angular ($\theta =D$ in PHENIX
notation) variables one 
gets according to Eq. (\ref{cos}) the nuclear index of refraction. Its value 
is found to be quite large $n=3$ compared to usual electromagnetic values 
close to 1. If interpreted in terms of the Breit-Wigner resonances, as 
explained below, it results in the large density of partons in the created   
quark-gluon system with about 20 partons within the volume of a single 
nucleon \cite{7}. It agrees with its estimates from $v_2$ and hydrodynamics.
This value is also related to the energy loss of gluons 
estimated in \cite{7} as $dE/dx\approx 1$ GeV/Fm. The height of the peaks 
determines the width of the ring which in its turn defines the free path 
length of Cherenkov gluons \cite{7} which happens to be long enough 
$R_f\sim 7$ Fm. Thus they hadronize, probably, close to the surface of the 
initial volume.

These estimates were obtained \cite{7} using the relation of the index of 
refraction to the forward scattering amplitude \cite{gw}
\begin{equation} 
{\rm Re} n(E )=1+\Delta n_R =1+\frac {6m_{\pi }^3\nu }{E^2}{\rm Re }F(E) =
1+\frac {3m_{\pi }^3\nu }{4\pi E } \sigma (E )\rho (E ). \label{n} 
\end{equation} 
Here $E$ denotes the energy, $\nu $ is the number of scatterers within 
a single nucleon, $m_{\pi }$ the pion mass, $\sigma (E)$ the cross section and 
$\rho (E)$ the ratio of real to imaginary parts of the forward scattering 
amplitude $F(E)$. Thus the emission of Cherenkov gluons is possible only for 
processes with positive ${\rm Re} F(E)$ or $\rho (E)$. It is well known that 
this requirement is fulfilled within one of the wings of any Breit-Wigner 
resonance\footnote{At the maximum of the resonance ${\rm Re } F(E)=0$ as seen 
below in Eq. (\ref{BW}). In particular, this is used to solve the problem of 
abundance of elements in the Universe (e.g. see \cite{fey}).}. Gluons with 
wide energy  spectrum are emitted during the collision. However only those 
whose energy fits the corresponding wing of the resonance (e.g., $\rho $-meson) 
which they form with the thermalized (or any other) gluons of the medium 
during the hadronization process can satisfy the requirement $n>1$. Inserting 
the Breit-Wigner shapes in Eq. (\ref{n}) one gets for a single resonance
\begin{equation} 
{\rm Re} n(E)=1+\frac {2J+1}{(2s_1+1)(2s_2+1)}\frac {6m_{\pi }^3\Gamma _R 
\nu } 
{EE_R^2}\frac {E_R-E}{(E-E_R)^2+\Gamma _R^2/4}. \label{BW} 
\end{equation} 
Here $J,\, s_1,\, s_2$ are spins of the resonance and its decay products, 
$E_R,\, \Gamma _R$ are its position and width. The above estimates of parton 
density $\nu $ follow from this expression for the nuclear index of refraction
with account of all mesonic resonances (sum over $R$ in Eq. (\ref{BW})). 
It also predicts the unusual particle content within the ring because only
energies $E<E_R$ matter to get $n>1$.

{\bf Prediction 2.} According to Eq. (\ref{BW}) the resonances formed within
the ring have masses shifted to smaller values somewhat below $m_R-\Gamma _R/2$ 
with asymmetrical distribution when reconstructed from their decay products. 

{\bf Proposal 2.} Plot the distribution of masses of $\pi ^+\pi ^-$ or 
$e^+e^-$-pairs in the ring. 

In the $\rho $-meson region it will be peaked slightly below $m_{\rho }-
\Gamma _{\rho }/2=700$ MeV if no shift is added due to the medium. The 
attenuation of Cherenkov gluons is moderate at these masses as can be shown 
from Eq. (\ref{BW}) (see \cite{hang}) and follows from above values of $dE/dx$ 
and $R_f$. The $e^+e^-$-mode is less probable but has much lower background. 
Particles momenta are relativistic \cite{hang}.

Let us stress that we do not require $\rho $-mesons or other resonances 
pre-exist in the medium but imply that they are the modes of its excitation 
formed during the hadronization process of partons. 
The Cherenkov gluon emission is a collective response of the quark-gluon medium 
to impinging partons related to its hadronization properties. It is determined 
by the energy behavior of the second term in Eq. (\ref{BW}). 

For the sake of simplicity, Eqs. (\ref{n}) and (\ref{BW}) valid at small 
$\Delta n_R$ typical for gases are used here. The value $n=3$ corresponds to a 
dense liquid. Therefore, it is proper to use the formula \cite{fey}
\begin{equation}
\frac {n^2-1}{n^2+2}=\frac {m_{\pi }^3\nu \alpha }{4\pi }=\sum _R
\frac {2J_R+1}{(2s_1^R+1)(2s_2^R+1)}\cdot \frac {4m_{\pi }^3\Gamma _R \nu }
{EE_R^2}\cdot \frac {E_R-E}{(E-E_R)^2+\Gamma _R^2/4},   \label{liq}
\end{equation}
where $\alpha $ denotes the colour polarizability of the colour-neutral 
medium. The value $\nu $ obtained from this expression is almost twice lower 
than given above. It does not change the qualitative conclusions about the 
dense medium (for more details see \cite{hang}). From Eq. (\ref{liq}) one can 
easily estimate that ${\rm Re} n$ is more than three times larger ${\rm Im} n$
at the maximum of the shifted resonance.

The Cherenkov gluons discussed above are comparatively low energy ones and coalesce 
to resonances. They originate from those regions of positive real part of 
the forward scattering amplitude which are bound within the left wings of
the resonances. 
However, from dispersion relation predictions and experiments with various 
colliding hadrons \cite{dnaz, blo} we know that there exists the high energy 
region of hadronic reactions where the real part of the forward scattering 
amplitude (or $\rho (E)$ ) is positive for 
all colliding partners. It happens at energy exceeding $E_{th}$=70 - 100 
GeV in the target rest system. Considering it as a common property of hadron 
reactions, we hope that high energy gluons possess the similar feature as 
carriers of strong forces. 

{\bf Prediction 3}. The very high energy forward moving partons can emit 
high energy Cherenkov gluons producing jets.

{\bf Proposal 3}. Plot the pseudorapidity distribution of dense groups of
particles in individual events (now again with collision axis chosen as 
$z$-axis) and look for maxima at angles determined by Eq. (\ref{cos}).

There are no gluons with such energy at RHIC but they will become available 
at LHC. Namely such gluons were discussed in \cite{1, 2} in connection with the 
cosmic ray event at energy $10^{16}$ eV (in the target rest system $E_t$) with 
the ring-like structure first observed \cite{8}. This energy just corresponds 
to LHC energies. The partons emitting such gluons move with high energy in 
the forward direction. 
With ${\rm Re} F(E_t)$ fitted to experimental data and dispersion relation 
predictions at high energies one can expect (see \cite{1, 2}) that the excess
of $n$ over 1 behaves as
\begin{equation}
\Delta n_R(E_t)\approx \frac {a\nu_h}{E_t}\theta (E_t-E_{th}).   \label{nh}
\end{equation}
Here, $a\approx 2\cdot 10^{-3}$ GeV is a parameter of ${\rm Re} F(E_t)$ obtained
from experiment (with dispersion relations used) and $\nu _h$ is the parton 
density for high energy region. It 
can differ from $\nu $ used at low energies. $\Delta n_R(E_t)$ decreases with 
energy for constant $\nu _h$. In this case Eq. (\ref{n}) should be applicable.
It would imply that the medium reminds a gas but not a liquid for very high 
energy gluons, i.e. it becomes more transparent.

The angles of the cone emission in c.m.s. of LHC experiments must be very large 
nevertheless (first estimates in \cite{2} are 60$^o$ - 70$^o$), i.e. 
the peaks can be seen in the pionization region at central pseudorapidities. In 
more detail it is discussed in \cite{1,2,hang,ag}. In this region the background 
is large, and some methods to separate the 
particles in the cone from the background were proposed in \cite{ag, 9}. 

The main difference between the trigger experiments at RHIC and this 
nontrigger experiment is in the treatment of the rest system of the medium. 
The influence of the medium motion on cone angles was considered in \cite 
{ssm}. It is important because all the above formulas are valid for emission 
in the rest system of the medium. 

At RHIC, the 90$^o$ trigger jet defines 
the direction of the away-side jet. Because of position of the trigger 
perpendicular to the collision axis of initial ions, the accompanying 
partons (particles) feel the medium at rest on the average in the c.m.s.
The similar trigger experiments are possible at LHC.
It is important to measure the cone angles for different angular 
positions of the trigger to register the medium motion. 

However, in nontrigger experiments with forward 
moving high energy partons inside of one of the colliding ions, the rest 
system of the medium is the rest system of another colliding ion. 
Therefore the cone angle should be calculated at that system and then 
transformed to the c.m.s. That is why these angles are so large even at 
small values of the refractivity index for high energy gluons. The low energy
gluons are unobservable here because they fly backward inside the accelerator 
pipe (about 180$^o$ in c.m.s.).

According to experimental data and dispersion relations predictions for
hadronic reactions, there exists wide energy region below $E_{th}$
where the real part of the 
forward scattering amplitudes is negative. Cherenkov gluons can not be emitted 
in this region. The transition gluon radiation can, nevertheless, happen.
For small $\Delta n_R$ it is proportional to $\Delta n_R^2$ and negligible but
can become important for comparatively large  $\Delta n_R$ like those
observed in the resonance region.

The analogy of gluons to photons is fruitful but colour polarizability asks 
for further investigation. The hadronic index of refraction can, in principle,
be determined from the gluon polarization operator in the strong gluon field
which is unknown yet. The role of the finite size of the nuclear medium was 
considered in \cite{2}. The parameter defining the notion of finiteness was 
formulated. It is important for small indices of the refraction.
The energy dependence of parton density must be studied in experiment. 
Search for Cherenkov gluons in other hadronic reactions (see \cite{8, ag}) 
is necessary.

This work was supported by RFBR grants 05-02-39028, 06-02-17051.

\end{document}